\documentclass[showpacs,twocolumn,prb,aps,superscriptaddress]{revtex4}
\usepackage{graphicx}
\begin{document}

\title{Phonon softening and ``forbidden'' mode observed by Raman scattering in Na$_{0.5}$CoO$_2$}
\author{Qingming Zhang}
\email{qmzhang@ruc.edu.cn} \affiliation{Department of Physics,
Renmin University of China, Beijing 100872, P. R. China}
\affiliation{National Laboratory of Solid State Microstructures,
Department of Physics, Nanjing University, Nanjing 210093, P. R.
China}
\author{Ming An}
\author{Shikui Yuan}
\author{Yong Wu}
\affiliation{National Laboratory of Solid State
Microstructures, Department of Physics, Nanjing University,
Nanjing 210093, P. R. China}
\author{Dong Wu}
\author{Jianlin Luo}
\author{Nanlin Wang}
\affiliation{Beijing National Laboratory for Condensed Matter
Physics, Institute of Physics, Chinese Academy of Sciences,
Beijing 100080, P. R. China}
\author{Wei Bao}
\affiliation{Los Alamos National Laboratory, Los Alamos, New
Mexico 87545, USA}
\author{Yening Wang}
\affiliation{National Laboratory of Solid State
Microstructures, Department of Physics, Nanjing University,
Nanjing 210093, P. R. China}
\date{\today}

\begin{abstract}
Polarized Raman scattering measurements have been performed on
Na$_{0.5}$CoO$_2$ single crystal from 8 to 305 K. Both the
$A_{1g}$ and $E_{1g}$ phonon modes show a softening below
$T_{c1}\approx 83$ K. Additionally, the
$A_{1g}$ phonon mode, which is forbidden in the scattering
geometry of cross polarization for the triangular CoO$_2$ layers,
appears below $T_{c1}$. In contrast, the metal-insulator
transition at $T_{c2}\approx 46$ K has only secondary effect on
the Raman spectra. The phonon softening and the ``forbidden'' Raman
intensity follow closely magnetic order parameter and the gap
function at the Fermi surface, indicating that the distortion
of CoO$_6$ octahedra at $T_{c1}$, instead of the Na ordering
at $\sim 350$ K, is the relevant structural component of the 83 K
phase transition.

\end{abstract}


\maketitle

\section{Introduction}

The layered cobaltates Na$_x$CoO$_2$ are well known for their
ionic mobility\cite{Molenda}. Recently, their
electronic properties have attracted much interest due to the
discovery of superconductivity when water is
intercalated\cite{Takada}. The Co ions form a triangular lattice
in the CoO$_2$ planes, which are separated by Na$^+$ ions. With
varying Na$^+$ content $x$, the valence of Co ions can be tuned
from Co$^{4+}$ (low spin $S=1/2$) to Co$^{3+}$ ($S=0$). Analogy to
the high transition-temperature cuprate superconductors has been
drawn, since Na$_x$CoO$_2$ may be regarded as doping a $S=1/2$
Mott insulator on a triangular lattice\cite{Takada}. However, the
Na$^+$ layers serve not only as charge reservoirs. Order of Na$^+$
ions at specific concentrations dramatically affect electronic
properties, such as the insulating state at $x$=0.5 sandwiched by
metallic states at lower and higher dopings\cite{Foo,Zandbergen}.
Additionally, the valence of Co ions in the hydrated
superconductor Na$_{0.35}$CoO$_2\cdot 1.3$H$_2$O is suggested in
recent studies to be close to +3.5 rather than the apparent +3.7,
caused by the isovalent exchange of the hydronium ion H$_3$O$^+$
and Na$^+$\cite{Barnes,Sakurai}. Hence, Na$_x$CoO$_2$ with $x$=0.5
instead of $x\sim0$ may be regarded as the parent compound of the
hydrated superconductor. Therefore, it is of particular interest to focus on
Na$_{0.5}$CoO$_2$.

A metal-insulator transition (MIT) at $T_{c2}\sim 51$ K is clearly
indicated in transport and infrared measurements of
Na$_{0.5}$CoO$_2$\cite{Foo,NLWang}. The transition is marked also
by a depression in magnetic susceptibility. In addition, another
anomaly in magnetic susceptibility occurs at $T_{c1}\sim 88$ K,
where the spins form an alternating antiferromagnetic pattern in
the CoO$_2$ plane\cite{Gasparovic,yokoi} and the Hall coefficient
abruptly changes sign\cite{Foo}. The MIT has been explained as a
consequence of charge ordering\cite{Foo}, and structure data from electron
and neutron diffraction measurements have been explained by
alternating Co$^{3.5+\delta}$ and Co$^{3.5-\delta}$ chains,
decorated by an ordered Na$^+$ pattern which exists already at the
room temperature\cite{Zandbergen,huang}. However, the charge
segregation into Co$^{3.5\pm\delta}$ is
disputed by a recent NMR study, and an alternative explanation of
nesting of the Fermi surface at the lower orthorhombic symmetry is
provided\cite{Bobroff}. The nesting of the Fermi surface is
also used to explain the angle-resolved photoemission spectroscopy (ARPES)
experiments\cite{Qian}. The subtle relation between electronic,
magnetic and structural properties also manifests in the restoration of the
triangular symmetry of Na$_{0.5}$CoO$_2$ at high magnetic
fields\cite{Balicas}.

Although a link between Na$^+$ ordering, which breaks the triangular symmetry, and the phase-transitions at $T_{c1}$ and $T_{c2}$
in Na$_{0.5}$CoO$_2$ has been suggested through Co charge segregation,
the relation between structure and the successive transitions can not be unambiguously determined in the crystallography studies due to the limited sensitivity of the diffraction techniques\cite{Zandbergen,huang}. In addition, no anomaly in lattice dynamics has been reported at these transitions in Na$_{0.5}$CoO$_2$.
Here we present polarized Raman scattering measurements on
single crystalline sample of Na$_{0.5}$CoO$_2$ from 8 to 305 K.
Phonon softening is observed below $T_{c1}$ for both the $A_{1g}$ and $E_{1g}$ modes of the CoO$_6$ octahedra.
Simultaneously, the forbidden $A_{1g}$ mode in the channel of cross polarization appears at $T_{c1}$.
It indicates that the transition at $T_{c1}$ is not only magnetic or
electronic, but also involved with structural distortion of the CoO$_6$ octahedra. The distortion in the CoO$_2$ plane, instead of the orthorhombic Na
ordering at $\sim 350$ K\cite{Foo,Zandbergen}, is more directly related to
the phase transition probed in the NMR, ARPES and neutron diffraction experiments at $T_{c1}$.

\section{experimental details}

Single crystal samples of Na$_{x}$CoO$_2$ were grown using a
traveling solvent floating-zone furnace.
The as-grown crystal was cut into pieces with the
dimensions of $\sim$$2\times 2\times 0.2$ mm$^3$ and
has a starting sodium concentration of about 0.75.
After the chemical deintercalation of
Na in solutions of I$_2$ dissolved acetonitrile, the actual Na
concentration was determined to be $x$=0.5$\pm$1\%
by inductively coupled plasma spectrometry.
The detailed procedure for preparing high-quality single crystal samples
can be found elsewhere\cite{Wu}. The susceptibility and
resistivity of the Na$_{0.5}$CoO$_2$ single crystal used in
the Raman study are shown in Fig.~1.
They are in good agreement with published results\cite{Foo,NLWang,Gasparovic,yokoi,Bobroff,Qian,huang},
with $T_{c2}\approx 46$ K and $T_{c1}\approx 83$ K.
\begin{figure}[tb]
\includegraphics[width=7.7cm,angle=0]{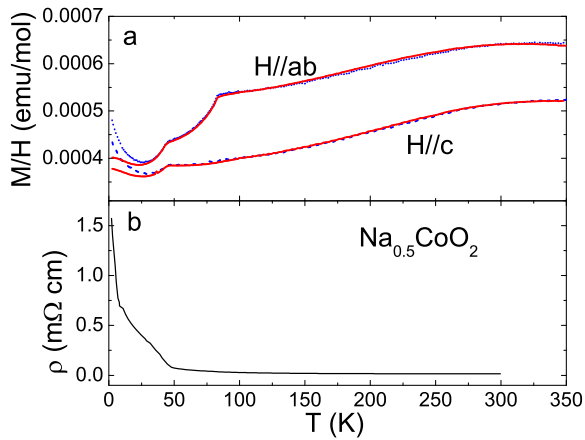}
\caption{(color online) Temperature dependence of (a) magnetic
susceptibility and (b) in-plane resistivity of Na$_{0.5}$CoO$_2$.
The blue dashed and red solid curves in (a) were measured in a
magnetic field of 1 and 12 T, respectively. } \label{fig1}
\end{figure}

The Raman measurements were performed with a double-gratings monochromator
(Jobin Yvon U1000). The detector is a back-illuminated CCD cooled by
liquid nitrogen. An argon ion laser was used with an excitation
wavelength of 514.5 nm. The laser beam of 3 mW was focused into a spot
of $\sim$70 microns in the diameter on the sample surface. The temperature
increase by laser heating is less than 10 K and was calibrated in
the measurements. The sample was mounted in a liquid helium
cryostat with a vacuum of $\sim$10$^{-7}$ torr. The data was collected
with a pseudo-backscattering configuration. Perfect
surface was obtained after cleavage, and the crystal orientation was
determined by the Laue pattern before the sample was mounted in the
cryostat.

\section{results and discussions}
The edge-shared CoO$_6$ octahedra in Na$_{x}$CoO$_2$ form triangular planes\cite{huang}. The polarization of light along the Co-O bond orientation is denoted as x, and the perpendicular orientation in the CoO$_2$ plane as y.
A standard calculation of the Raman tensors shows
that the in-plane $E_{1g}$ mode of oxygen vibrations in the CoO$_6$
octahedra is allowed in both the xx and xy channels.
The out-of-plane $A_{1g}$ mode, however, is allowed only in the xx channel.  Polarized Raman spectra in the xx and xy channels, measured at
various temperatures, are shown in Fig.~2.
The wavenumbers of the $E_{1g}$ and $A_{1g}$ modes are 469 cm$^{-1}$
and 572 cm$^{-1}$, respectively, at
305 K. They are consistent with published data on Na$_{x}$CoO$_2$
and Na$_{x}$CoO$_2\cdot y$H$_2$O\cite{Iliev,Qu,Lemmens} and
first-principles calculations\cite{LiZY,Zhang}.
\begin{figure}[tb]
\includegraphics[width=4.2cm]{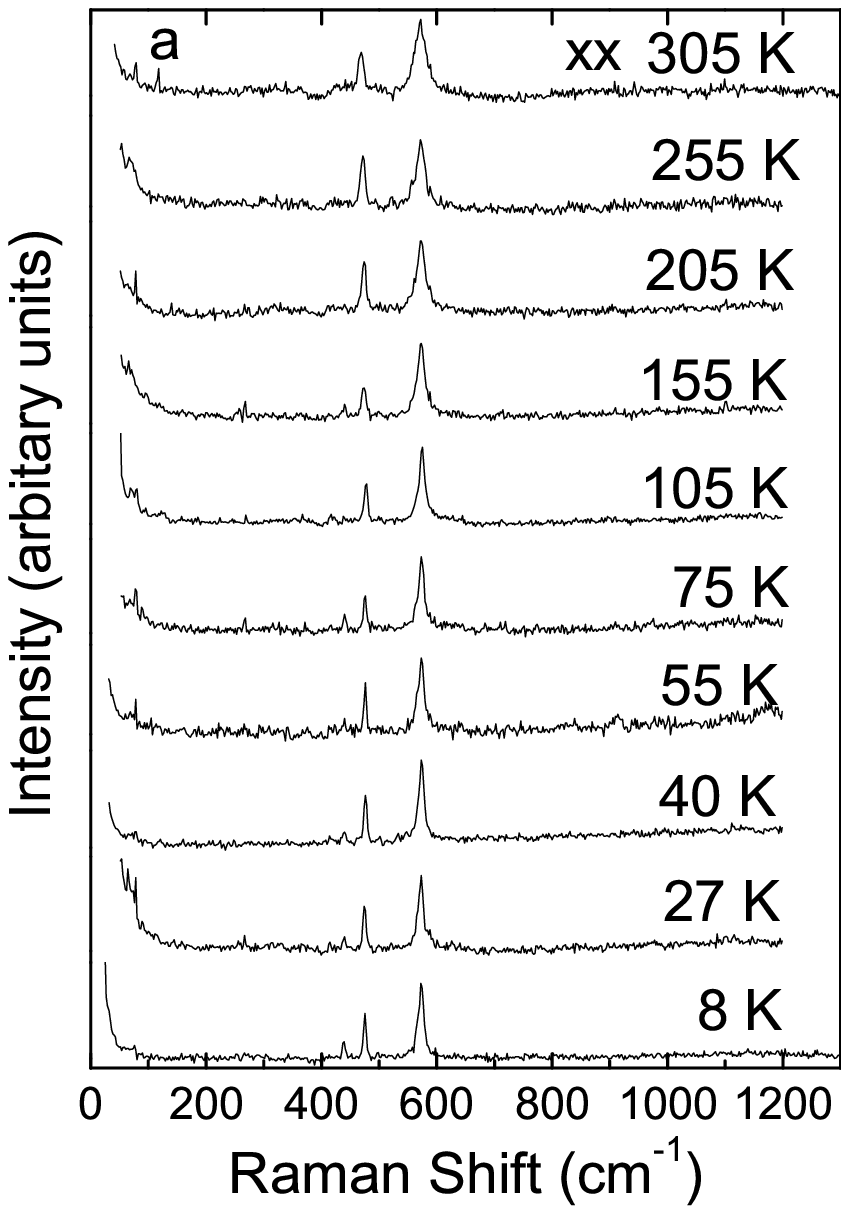}
\includegraphics[width=4cm]{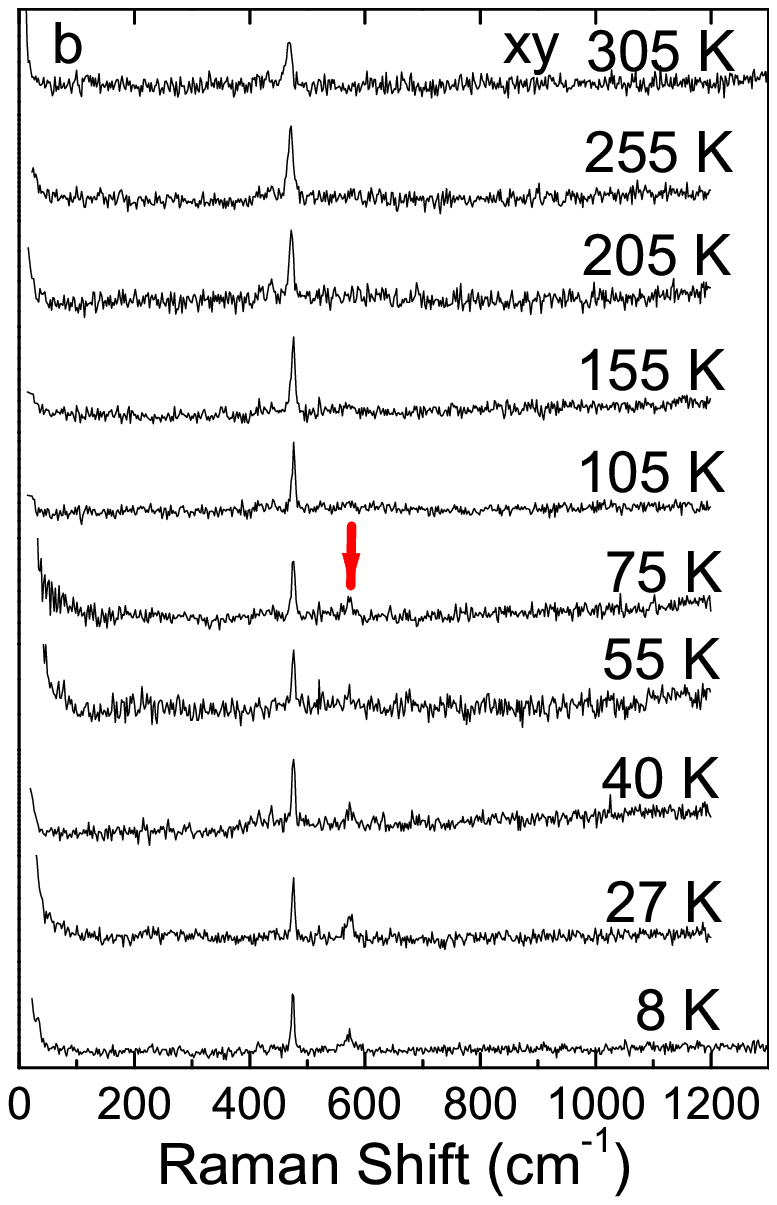}
\caption{(color online) Polarized Raman spectra at various
temperatures. (a) In the xx channel, both $E_{1g}$ and $A_{1g}$
modes are allowed. (b) In the xy channel, only the $E_{1g}$ mode is
active. However, the ``forbidden'' $A_{1g}$ mode near 573
$cm^{-1}$ grows in the xy channel below $T_{c1}\approx 83$ K, as
indicated by the arrow.} \label{fig2}
\end{figure}

The wavenumbers of the $E_{1g}$ and $A_{1g}$ modes are shown in
Fig.~3 as a function of temperature. As the sample was cooled from
the room temperature to $T_{c1}$, the phonon wavenumbers of both
modes increase gradually. The increase can be understood naturally in
terms of anharmonic effect which has also been reported previously\cite{Lemmens}. It includes two contributions: one from
thermal expansion and the other from multi-phonon decay
process\cite{Cowley,Klemens}.
The contribution from thermal expansion can be written as
$-3\omega_0\gamma\cdot\Delta a/a$, where $\omega_0$ is the
zero-temperature phonon frequency, $\gamma$ the
Gr\"{u}neisen constant, and $\Delta a/a$ the linear thermal
expansion\cite{Menendez,Lang}. Neutron diffraction shows that
the average in-plane $\Delta a/a$ is 16 ppm, whereas $\Delta a/a$
is 2540 ppm for the c-axis\cite{Williams}.
Since the $E_{1g}$ ($A_{1g}$) phonon mode is related to the in-plane
(out-of-plane) vibrations of oxygen atoms in the CoO$_6$
octahedra, the wavenumbers should be sensitive to the change of
the in-plane bonds rather than the c-axis expansion. Hence,
the small in-plane $\Delta a/a$ would translate to a small
anharmonic contribution from thermal expansion.
\begin{figure}[tb]
\includegraphics[width=8cm,angle=0]{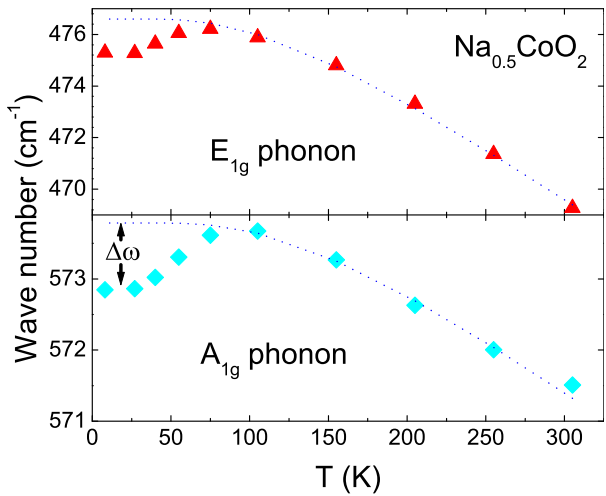}
\caption{(color online) Temperature dependence of the $E_{1g}$ and
$A_{1g}$ phonon wavenumbers. The dotted curves are the least-square
fitting of data taken above 83 K to Eq.~(1)
for anharmonic effect, and phonon softening $\Delta\omega$ occurs below $T_{c1}$.} \label{fig3}
\end{figure}

The main contribution to the shift of phonon wavenumbers above $T_{c1}$
then can be reasonably attributed to
multi-phonon decay process. For a three-phonon process, a
Raman optical phonon with the frequency of $\omega_0$ will decay
into two phonons with $\omega_1$ and $\omega_2$, respectively,
where $\omega_0=\omega_1+\omega_2$. The frequency shift by the process is
\begin{equation}\label{1}
\omega-\omega_0=-\alpha\cdot(\frac{1}{e^{\omega_1/kT}-1}+\frac{1}{e^{\omega_2/kT}-1}),
\end{equation}
where $\alpha$ is a material-dependent constant\cite{Cowley,LiuMS}.
Klemens has considered the decay
channels through acoustic phonons which fulfill the condition
$\omega_1=\omega_2$\cite{Klemens}. In many cases, it offers
a good approximation. The dotted curves in Fig.~3 are the results
of fitting data above $T_{c1}$ to Eq.~(1) under the Klemens
condition with $\alpha=7.5$ and 3.5 cm$^{-1}$ for the $E_{1g}$ and
$A_{1g}$ modes, respectively. The Klemens model describes well our
data above $T_{c1}$, and the values of $\alpha$ are comparable
with ~13 cm$^{-1}$ deduced from diamond, and 2 to 4 cm$^{-1}$ for
semiconducting materials such as AlN, Si, and GaAs\cite{LiuMS,
Song}.

When the sample was cooled further, the first new discovery of this work is that
both the $E_{1g}$ and $A_{1g}$ modes soften considerably below the $T_{c1}$,
departing from the dotted curves from the anharmonic effect (Fig.~3).
The softening in frequency, $\Delta \omega$, is taken as the
difference between the dotted curve and measured data. It is
scaled so that it has the same value at 8 K for both the
$E_{1g}$ and $A_{1g}$ modes, and it follows the same function
of temperature for both modes as shown in Fig.~4.

\begin{figure}[tb]
\includegraphics[width=8.2cm,angle=0]{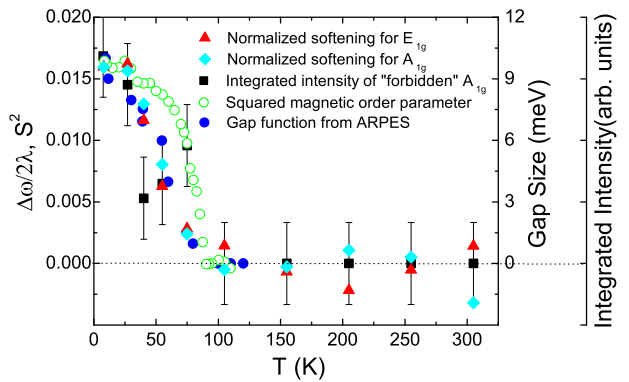}
\caption{(color online) Temperature dependence of phonon softening
$\Delta\omega$ for the $E_{1g}$ (red triangles) and $A_{1g}$
(cyan diamonds) modes, and the integrated intensity of the
``forbidden'' $A_{1g}$ mode in the xy channel (black squares). For
details on $\lambda$, see text. The open green circles represent
the squared magnetic order parameter $S^2$\cite{Gasparovic}, and
the filled blue circles the gap function at the Fermi
surface\cite{Qian}. } \label{fig4}
\end{figure}

Since an antiferromagnetic transition occurs at $T_{c1}$, it is
natural to consider the spin-phonon coupling. In a magnetic
material mediated by the superexchange interaction $J$, such as in
magnetic insulators, the $J$ would be modulated by
the vibrating positions of ligands:
\begin{equation}\label{2}
J(\mathbf{r})= J(\mathbf{r_0})+\left. \frac{\partial J}{\partial
\mathbf{r}}\right|_{\mathbf{r_0}} \cdot
(\mathbf{r}-\mathbf{r_0})+\left. \frac{\partial ^2 J}{\partial
\mathbf{r}^2}\right|_{\mathbf{r_0}} (\mathbf{r}-\mathbf{r_0})^2+\dots,
\end{equation}
where $\mathbf{r}$ and $\mathbf{r_0}$ are the instantaneous and
equilibrium positions of oxygen, respectively, in our case.
Usually spin-phonon coupling is a second-order effect since the
linear contribution is canceled out by the symmetric oscillation
of ions around equilibrium positions. However, some local
asymmetric structures would cause the linear part to be finite,
such as the buckled CuO$_2$ plane in
YBa$_2$Cu$_3$O$_{7-\delta}$\cite{Normand}. Similarly,
in Na$_{x}$CoO$_2$, the
oxygen and cobalt ions are not in the same plane. The
oscillation of oxygen ions is expected to have a strong effect on
magnetic interaction $J$.
The phonon softening due to the spin-phonon coupling can be expressed as\cite{Lockwood,Chen}
\begin{equation}\label{3}
\Delta\omega =-\lambda \langle
\mathbf{S_i}\cdot\mathbf{S_{i+1}}\rangle=2\lambda S^2,
\end{equation}
where $\lambda$ is the spin-phonon coupling
coefficient, $\langle\mathbf{S_i}
\cdot\mathbf{S_{i+1}}\rangle$ the average for adjacent
spin pairs, and $S^2=(M/2g\mu_B)^2$ the squared magnetic order parameter.
In evaluating $\langle\mathbf{S_i} \cdot\mathbf{S_{i+1}}\rangle$ in
the last step of Eq.~(3),
neutron diffraction result by G\v{a}sparovi\'{c} et al.\cite{Gasparovic} was used.
In Fig.~4, measured $S^2$ for Na$_{0.5}$CoO$_2$ (open green
circles)\cite{Gasparovic} is compared with the scaled phonon softening
for the $E_{1g}$ (red triangles) and $A_{1g}$ (cyan diamonds) modes. While
both $\Delta\omega$ and $S^2$ appear below $T_{c1}$, the softening
is slower and less mean-field-like than the squared magnetic order
parameter, namely, the Eq.~(3) is not perfectly followed. It is
not clear whether this is due to Na$_{0.5}$CoO$_2$ not being an
insulator above the 46 K metal-insulator transition. If one
nevertheless applies Eq.~(3) to Na$_{0.5}$CoO$_2$ at 8~K, which is
then an insulator, the spin-phonon coupling coefficient $\lambda$
is estimated as 30 cm$^{-1}$ for the $A_{1g}$ mode and 41
cm$^{-1}$ for the $E_{1g}$ mode, respectively. For comparison,
$\lambda$ is $-$50 cm$^{-1}$ in CuO\cite{Chen}, 6 and 9 cm$^{-1}$
for two phonon modes in Y$_2$Ru$_2$O$_7$\cite{Lee}, and
$|\lambda|$ is less than 3 cm$^{-1}$ for antiferromagnets in the
rutile structure such as FeF$_2$, MnF$_2$ and
NiF$_2$\cite{Lockwood}. Thus, the spin-phonon coupling in
Na$_{0.5}$CoO$_2$ is very strong.

The spin-phonon coupling is a {\em dynamical} spin-lattice
interaction. At a magnetic transition, {\em static} spin-lattice interaction
is also possible, which is usually referred to as
magnetoelastic effect. The effect requires no phonon softening,
but a static change of local structure.
In addition to the phonon softening, the second new discovery of this work is that
the $A_{1g}$ mode in
the ``forbidden'' xy channel also appears below $T_{c1}$ in
Na$_{0.5}$CoO$_2$, see Fig.~2(b).
The violation of the Raman selection rules mentioned
above for the triangular layers requires a structural distortion in the edge-shared CoO$_6$ to break lattice symmetry of the high temperature phase.
Qualitatively, the integrated intensity of the ``forbidden'' $A_{1g}$ mode
is proportional to the distortion at low temperatures.
In Fig.~4, the integrated intensity of the ``forbidden'' mode
is plotted as the black squares. It follows the general trend of
$\Delta\omega$, except that there appears a dip around $T_{c2}$.
However, more sensitive experiments are called for to
establish the statistical significance of the dip.
Magnon scattering was suggested as the origin of the ``forbidden'' $A_{1g}$ mode in view that
the mode appears only below the magnetic transition.
However, this explanation has difficulty for the same mode in the allowed
xx channel at all measured temperatures.

In Fig.~4, the single-particle gap function (filled blue circles) from
the ARPES measurements\cite{Qian} is also plotted. Different from
magnetic order parameter, it traces the phonon softening $\Delta\omega$
very well. On the other hand, the metal-insulator transition at
$T_{c2}$ has no detectable effect on either $\Delta\omega$ or the gap function. Although the gap is opened only at the $a_{1g}$ Fermi surface of
the Co ions, the Na ordering at $\sim 350$ K was
invoked to provide a nesting condition for the Fermi surface\cite{Qian}.
But Qian {\it et al.}\ also noticed that their observed gap is
less anisotropic than expected from such a nesting scenario and
that the gap size is rather soft compared to many
CDW systems. In addition, the Fermi surface geometry does not share
the two-fold symmetry of the Na supercell\cite{Qian}.
Thus, our observed structural transition directly involving
CoO$_6$ at the same temperature offers a promising alternative
mechanism for the band folding. Similarly, the supercell needed
for explanation of the NMR results at $T_{c1}$\cite{Bobroff} may
come instead from the
distortion of the CoO$_6$ octahedra. We notice that electron
diffraction experiments at 100 K detected from the Na layers
another tripled superlattice pattern which, however, disappears at
20 K\cite{Zandbergen,huang}.
Whether the tripled superlattice pattern has anything to do with our observed
CoO$_6$ distortion, which persists down to our lowest measurement temperature
at 8 K, is not obvious.

\section{conclusions}

In summary, our polarized Raman measurements of
Na$_{0.5}$CoO$_2$ single crystal have shed new light on the multiple
phase transitions in the interesting material. Above $T_{c1}\approx 83$ K,
only conventional anharmonic effect was observed. The softening of the
$A_{1g}$ and $E_{1g}$ optical phonon modes was observed
below $T_{c1}$, accompanied by the $A_{1g}$ mode in the
``forbidden'' xy Raman channel. Our results indicate a
structural transition at $T_{c1}$ which breaks the
local lattice symmetry of the CoO$_6$ octahedra. The spin-phonon
coupling in Na$_{0.5}$CoO$_2$ is among the strongest of magnetic materials.
The closely related order-parameters of the structural distortion and
phonon softening in the CoO$_6$ layers as well as the magnetic and
electronic transitions involving electrons of Co ions
suggests a common origin for these transitions. The structural
condition for the nesting Fermi surface scenario which was invoked
as the mechanism for the magnetic transition and the partial gapping
of the Fermi surface at $T_{c1}$ is more likely the structural
distortion discovered in this work in the Co layers than the
orthorhombic Na ordering at $\sim$350 K.
If Na$_{0.5}$CoO$_2$ is indeed more appropriate as the parent
compound of the superconducting Na$_{0.35}$CoO$_2\cdot 1.3$H$_2$O\cite{Barnes,Sakurai},
it may be difficult to exclude a lattice contribution to focus on
a purely electronic superconducting mechanism.

\section{acknowledgments}

The authors would like to thank B. Normand, T. Li, Q. H. Wang, Z.-Q.
Wang and Y. Chen for helpful discussions. The work was supported by
the MOST of China (973 project No:2006CB601002) and NSFC Grant No.
10574064; WB was supported by the US DOE.

\end{document}